\documentclass{webofc}
\usepackage[varg]{txfonts}   
\usepackage{xcolor}
\usepackage{subfigure}
\usepackage{graphicx}
\usepackage{caption}
\usepackage{float}
\usepackage{subcaption}
\usepackage{wrapfig}
\usepackage{lineno}
\begin{document}
%
\title{New Insights into Global Spin Alignment of Vector Mesons Using Relativistic Heavy-Ion Collisions from STAR}
%
%

\author{\firstname{Baoshan} \lastname{Xi}\inst{1}\fnsep\thanks{\email{xibaoshan@fudan.edu.cn}} for the STAR Collaboration}
\institute{Fudan University, 220 Handan Rd., Yangpu District, Shanghai(200433), CHINA 
          }

\abstract{%
In these proceedings, we present new measurements from STAR of the global spin alignment($\rho_{00}$) of $\phi$ mesons in Au+Au collisions at $\sqrt{s_{\mathrm{NN}}}$ = 19.6, 14.6 GeV using BES-II data and of $J/\psi$ mesons in isobar collisions ${ }_{44}^{96} \mathrm{Ru}+{ }_{44}^{96} \mathrm{Ru}$ and ${ }_{40}^{96} \mathrm{Zr}+{ }_{40}^{96} \mathrm{Zr}$ at $\sqrt{s_{\mathrm{NN}}}$ = 200 GeV. The energy-dependent spin alignment for $\phi$ mesons from BES-II data is consistent with that published from BES-I and showcases a significant improvement in precision. We performed a new measurement assessing the rapidity dependence of 
$\phi$ meson $\rho_{00}$, and our findings are consistent with theoretical predictions. The global spin alignment of $J/\psi$ mesons with respect to first order event plane aligns with a value of $1/3$ within the statistical error. Additionally, we discuss the procedure of measuring the global spin alignment of $\rho^0$ mesons at RHIC and provide the projection for errors.
}
\maketitle
\section{Introduction}
\label{sec-intro}

In non-central heavy-ion collisions, a substantial amount of angular momentum is generated when two nuclei pass each other at near-light speed. This angular momentum results in the formation of a vorticity field within the system. Through spin-orbital coupling within this vorticity field, quarks can become polarized along the direction of the angular momentum\cite{Liang:2004xn, Liang:2004ph}. This quark-level polarization along a global direction is eventually reflected in the polarization of hyperons or the spin alignment of vector mesons, both of which can be measured in experiments.


The global spin alignment of particles produced in heavy-ion collisions can provide valuable insights into the strong force field and the properties of the quark-gluon plasma. Recently, the STAR collaboration measured the global spin alignment of $\phi$ and $\mathrm{K}^{* 0}$ in Au+Au collisions using data from the first phase of the RHIC Beam Energy Scan program (BES-I) \cite{STAR:2022fan}. In these measurements, the global spin alignment ($\rho_{00}$) of $\mathrm{K}^{* 0}$ was found to be consistent with $1/3$, whereas the $\rho_{00}$ of $\phi$ significantly exceeded $1/3$. The large signal for $\phi$ meson cannot be explained by conventional mechanisms but might be attributed to the influence of vector meson force fields. The observed signal of this global spin alignment aligns with a model incorporating $\phi$ meson field coupling to the $s$ and $\bar{s}$ quarks\cite{Sheng:2019kmk}, offering new insights into strong interactions.

Our interpretation gains depth from the inclusion of BES-II data collected at collision energies of $\sqrt{s_{NN}}$ = 14.6 and 19.6 GeV. With significantly larger statistics and the incorporation of a new event plane detector (EPD), these measurements substantially enhance detection accuracy. Exploring the relationship between global spin alignment and rapidity could provide intriguing insights into the nuances of the fluctuating local strong force field, as emphasized in the theoretical prediction\cite{Sheng:2023urn}.

A notable aspect allowing the assessment of contributions from the fluctuating strong force field in theory is the composition of $\phi$ mesons, which are made up of $s\bar{s}$ quarks from the same flavor family. This uniqueness makes the global spin alignment measurement for $J/\psi$ particularly interesting, given that $J/\psi$ is composed of $c\bar{c}$ quarks, which also belong to the same flavor family. In contrast, the $\rho^0$ meson, unlike $\phi$ and $J/\psi$, comprises a mixture of quark flavors $(u\bar{u} - d\bar{d})/\sqrt{2}$. Due to its short lifetime, $\rho^0$ mesons are continuously destroyed and regenerated. These factors make it less optimal as a probe for local fluctuations in the strong force field. However, the global spin alignment of $\rho^0$ mesons holds significant implications for Chiral Magnetic Effect (CME) analyses\cite{Shen:2022gtl,Tang:2019pbl}.

These studies, incorporating more differential measurements and exploring vector mesons with various quark contents, are geared towards comprehending the influence of vector meson force fields on the evolution of matter produced in heavy-ion collisions. By introducing new particle species and its decay channels, researchers can gain insights into the complex effects of in-medium interactions and hadronization mechanisms on global spin alignment.

\section{Analysis method}
\label{sec-analy}
We reconstructed vector mesons and obtained the distribution of $\cos \theta^*$, where $\theta^*$ is angle between the polarization direction $\hat{L}$ and the momentum direction of a daughter particle in the rest frame of the parent vector meson. We use Eq.~(\ref{eq:equation1}) to fit the $\cos \theta^*$ distribution, where the upper sign is used for $\phi$ and $\rho^0$ meson, and the lower sign is used for $J/\psi$\cite{Faccioli:2010kd}.

\begin{equation}
\frac{d N}{d \cos \theta^*}=N_0 \times\left[\left(1\mp\rho_{00}^{obs}\right)\pm\left(3\rho_{00}^{obs}-1\right) \cos ^2 \theta^*\right]
\label{eq:equation1}
\end{equation}
where $N_0$ is the normalization factor, and $\rho_{00}^{obs}$ is the observed global spin alignment. From an experimental standpoint, $\rho_{00}$ is reflected in the directional tendency of emitted decaying daughters. When $\rho_{00}=1/3$, $\cos \theta^*$ is uniformly distributed, indicating no global spin alignment. Conversely, a deviation of $\rho_{00}$ from $1/3$ results in a non-uniform distribution of $\cos \theta^*$.

To ensure the accurate determination of the physical global spin alignment, it is crucial to apply additional corrections for detector effects. We calculate detector efficiency using embedding technique\cite{STAR:2009sxc}. Finite acceptance in pseudorapidity could introduce artificial signals, and a finite event plane resolution might smear the signal strength. Both aspects have been corrected as outlined in\cite{Tang:2018qtu}. To validate the effectiveness of our correction method, we conducted simulation experiments considering limited detector acceptance and event plane resolution. The final corrected $\rho_{00}$ value exhibited excellent agreement with the initial input $\rho_{00}$ value.

\section{Results}
\label{sec-res}
\begin{figure}[h]
\vspace{-0.2cm}
\centering
\begin{minipage}{0.35\linewidth}
\vspace{-0.35cm}
\centering
\includegraphics[width=0.86\linewidth]{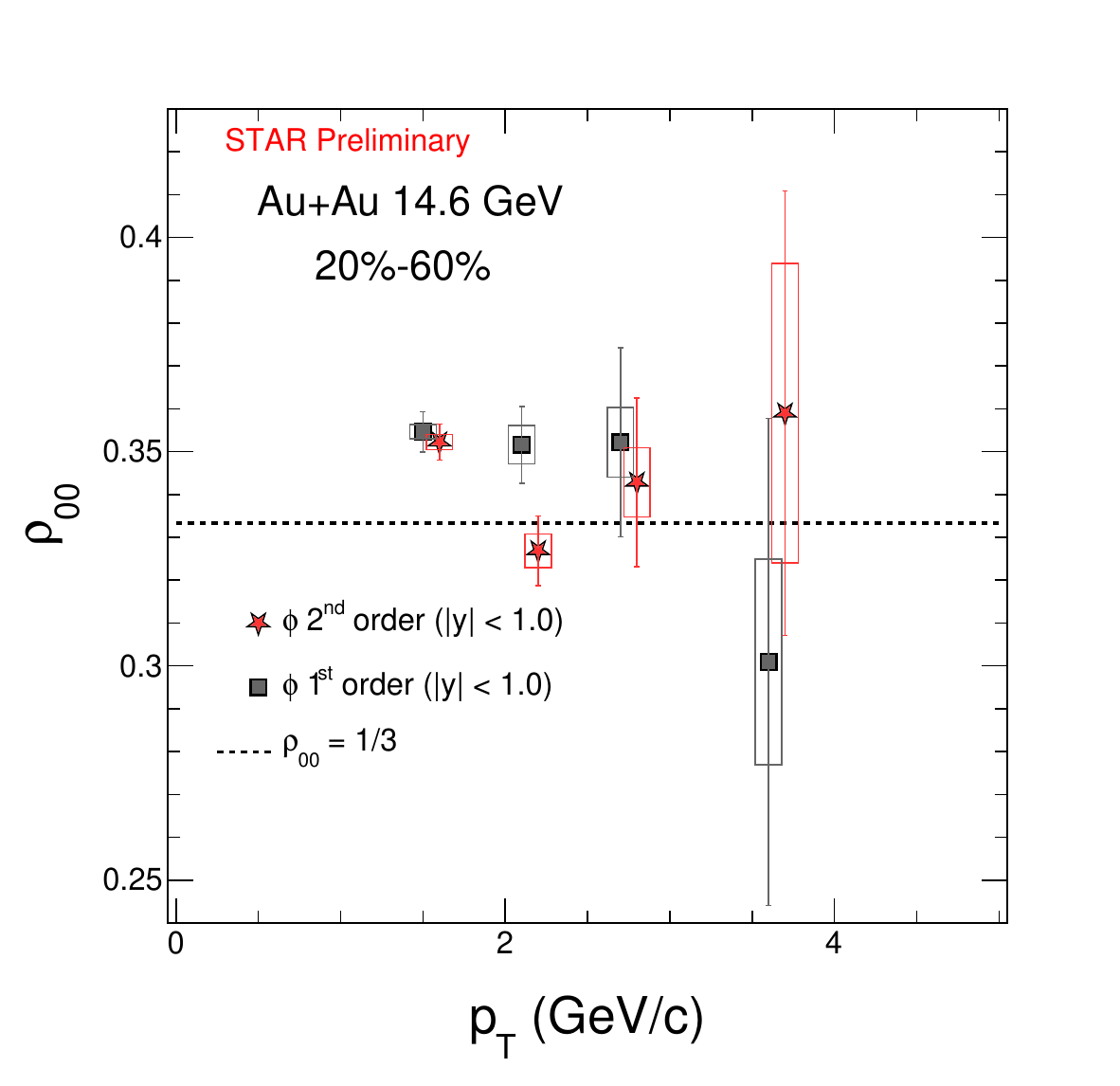}
\label{fig-1-a}
\end{minipage}
\begin{minipage}{0.305\linewidth}
\centering
\includegraphics[width=1.\linewidth]{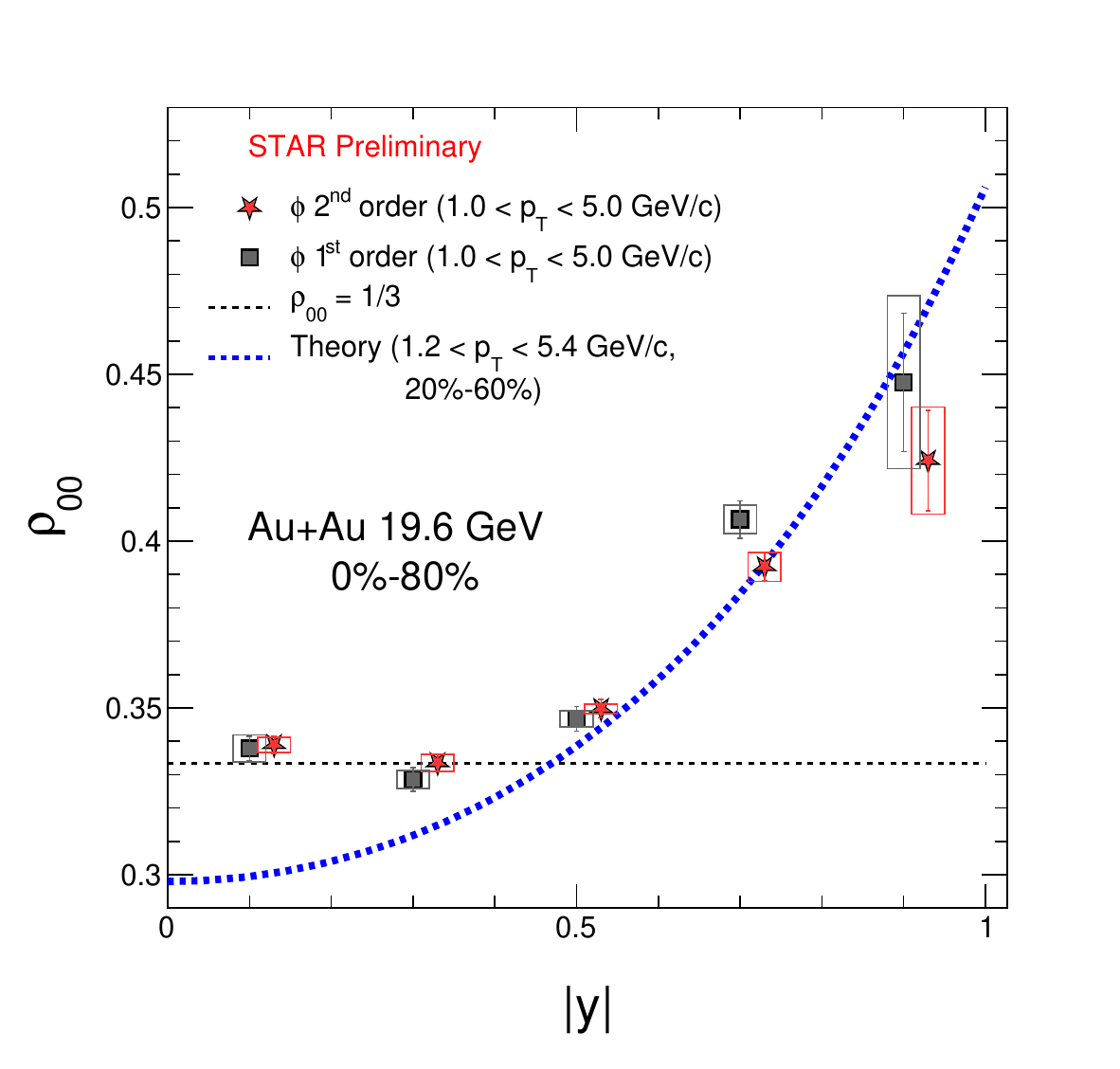}
\label{fig-1-a}
\end{minipage}
\vspace{-0.5cm}
\caption{Left panel : global spin alignment of $\phi$ as a function of $p_T$ for Au+Au collisions at $\sqrt{s_{\mathrm{NN}}}$ = 14.6 GeV, obtained with BES-II data. Right panel : rapidity dependence of global spin alignment for Au+Au collisions at $\sqrt{s_{\mathrm{NN}}}$ = 19.6 GeV. For both panels, the black and red symbols represent results relative to first order and second order EPs respectively.
}
\label{fig-1}
\end{figure}

The dependence of $\phi$'s global spin alignment on the transverse momentum $p_T$, was investigated in Au+Au collisions at $\sqrt{s_{\mathrm{NN}}}$ = 14.6 GeV during BES-II, as illustrated in the left panel of Figure~\ref{fig-1}. In the figure, black symbols denote results from measurements relative to the first order event plane, while red symbols indicate results relative to the second order event plane. The first order EP gives better correlation to direction of $\hat{L}$, while second order EP gives better precision results. Notably, there is no significant difference observed in the measured $\rho_{00}$ values within uncertainties. The BES-II results substantially reduced measurement errors and are consistent with those from BES-I\cite{STAR:2022fan}.

A new study examining the rapidity dependence of global spin alignment was conducted using BES-II data. The right panel of Figure~\ref{fig-1} displays the results from Au+Au collisions at 19.6 GeV. In the figure, black symbols results relative to first order EP and red symbols relative to second order EP. All results exhibit significant rapidity dependence. The theoretical prediction for rapidity dependence at 19.6 GeV displays similar trend in data for $|y|$ > 0.5, depicted by the blue curve in the right figure\cite{Sheng:2023urn,Sheng:2022wsy}. This observed phenomenon may be attributed to larger fluctuations of the strong force field in the meson's rest frame enhanced by its motion in directions perpendicular to that of global angular momentum\cite{Sheng:2023urn}.

\begin{wrapfigure}{l}{0.4\textwidth}
\vspace{-0.9cm}
  \begin{center}
    \includegraphics[width=0.4\textwidth]{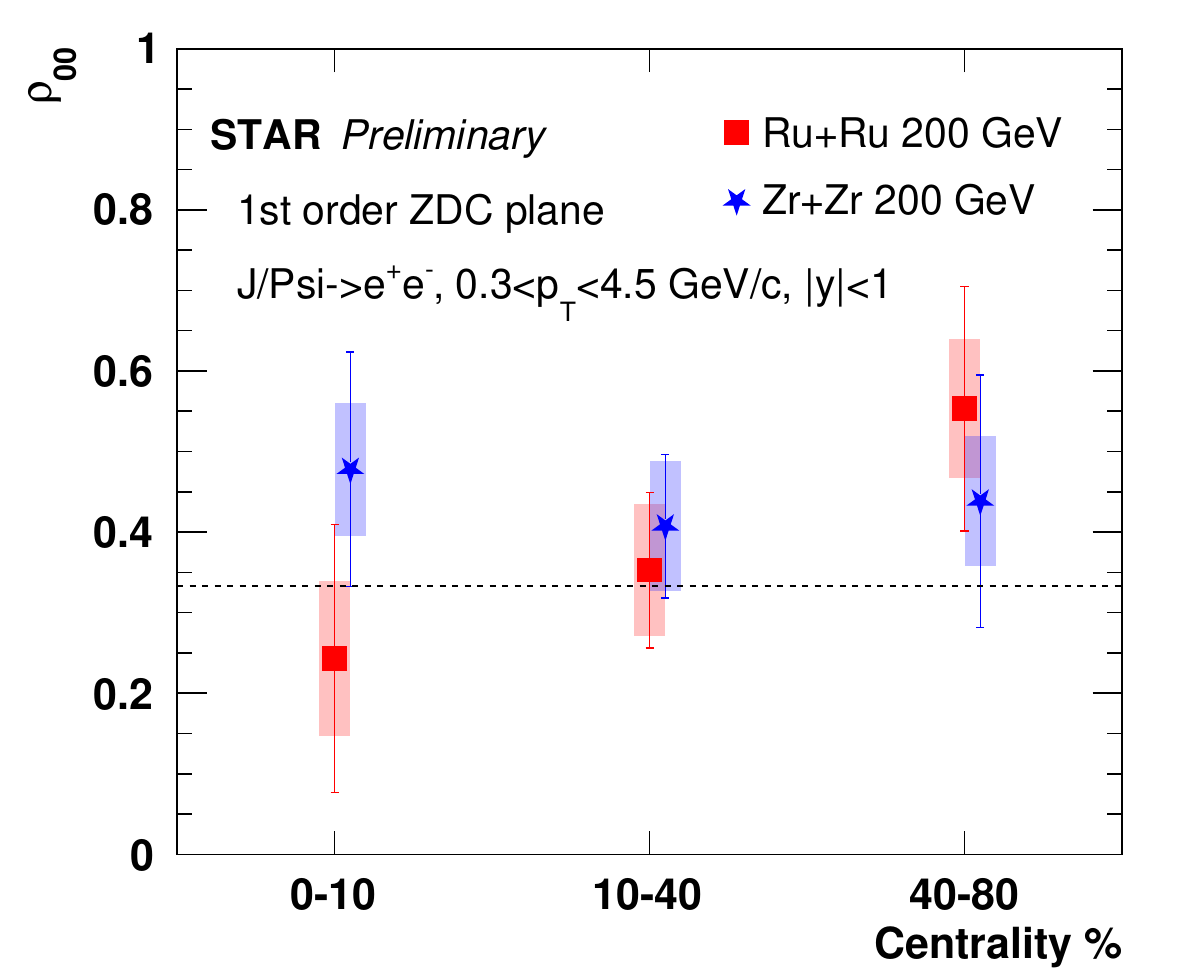}
  \end{center}
  \vspace{-0.7cm}
  \caption{Global spin alignment of $J/\psi$. The red and blue symbols represent the result of Ru+Ru and Zr+Zr, respectively.}
\vspace{-1.0cm}
\label{fig-2}
\end{wrapfigure}

In Figure~\ref{fig-2}, we present the global spin alignment measurements of $J/\psi$ using the first-order event plane in isobar collisions at $\sqrt{s_{\mathrm{NN}}}$ = 200 GeV. The red symbols represent the results for Ru+Ru collisions, while the blue symbols represent the results for Zr+Zr collisions. Across central to peripheral collisions, the data are consistent with $1/3$ within the margin of error. Notably, the ALICE collaboration observed that signals of global spin alignment of $J/\psi$ are smaller than $1/3$ in PbPb collisions\cite{ALICE:2022dyy}, while the global spin alignment of $D^*$ at high $p_T$ is greater than $1/3$\cite{Micheletti}. These findings provide additional insights for theorists to unravel the underlying mechanisms governing global spin alignment induced by the strong force field.


\begin{figure}[h]
\vspace{-0.3cm}
\centering
\sidecaption
\includegraphics[width=5.5cm,clip]{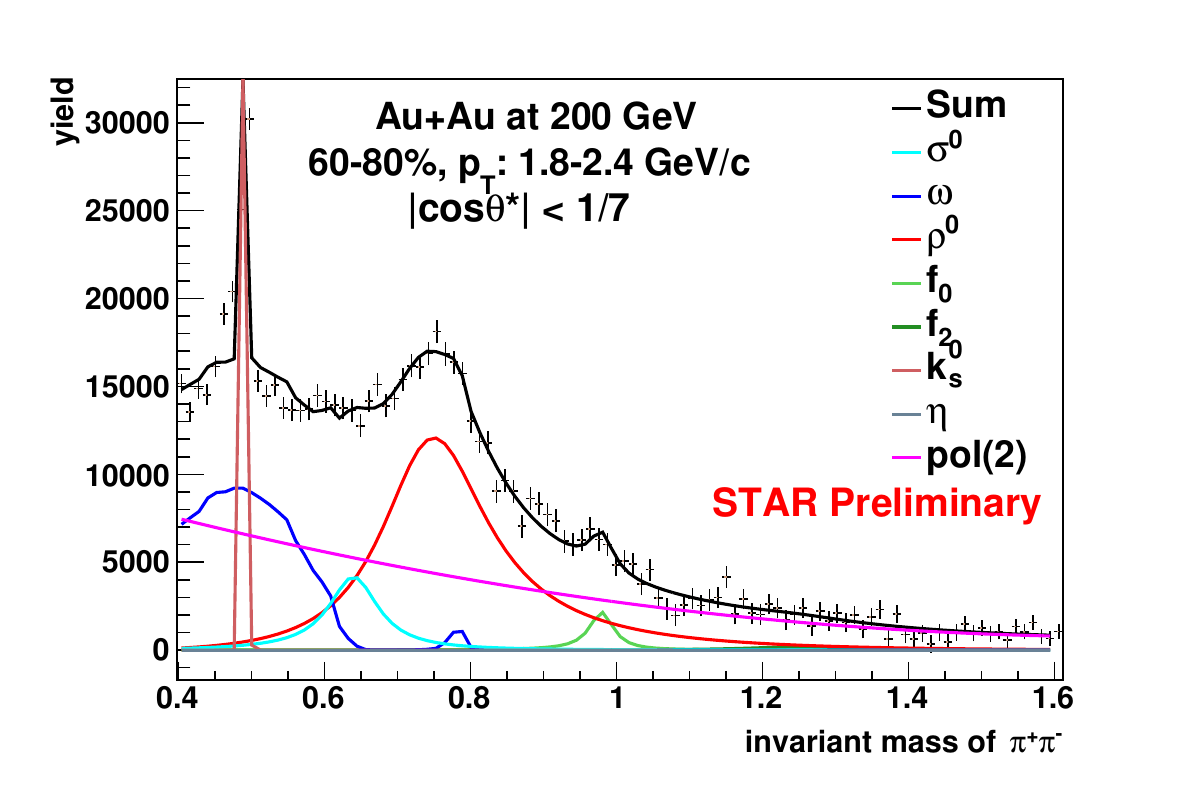}
\caption{Cocktail fitting of yield in a $\cos \theta^*$ bin for $\rho^0$ meson, where the red curve represents the signal of $\rho^0$. The plot is for $p_T$ within 1.8-2.4 GeV/c and centrality within 60-80\% in Au+Au collisions at $\sqrt{s_{\mathrm{NN}}}$ = 200 GeV.}
\label{fig-6 }       
\vspace{-0.6cm}
\label{fig-3}
\end{figure}

In a recent phenomenological study, the relation between the global spin alignment of $\rho^0$ mesons and CME observable $\Delta \gamma_{112}$ was explored\cite{Shen:2022gtl,Tang:2019pbl}. Conversely, the influence of $\rho^0$ meson $\rho_{00}$ on the CME-background-control method, such as the spectator/participant plane ratio of the CME observables or the event shape engineering, requires further dedicated investigation. The study\cite{Shen:2022gtl,Tang:2019pbl} revealed a clear dependence of CME observables on $\rho^0$ meson's $\rho_{00}$. Investigating the global spin alignment of $\rho^0$ mesons is crucial to understanding impact on CME observables. However, extracting $\rho^0$ meson's $\rho_{00}$ is complicated by the cocktail background in the invariant mass distribution. We reconstructed the invariant mass spectrum through the decay channel $\rho^0 \rightarrow \pi^{+} \pi^{-}$ with Au+Au and isobar collisions at $\sqrt{s_{\mathrm{NN}}}$ = 200 GeV. The spectrum involves contributions from 7 particles: $\omega, \rho^0, f_0, f_2, \sigma^0, k_s^0$ and $\eta$. This complexity makes the extraction of $\rho^0$ yield highly challenging. In Figure~\ref{fig-3}, We present the cocktail fitting results of a $\cos \theta^*$ bin for $\rho^0$ with $p_T$ of 1.8-2.4 GeV/c in Au+Au collisions. This study is currently in progress. Currently, our estimation suggests that the absolute uncertainty of $\rho_{00}$ that can be obtained from combined Au+Au collision data in 2023 and 2025 is on the order of 0.01.

\section{Summary}
\label{sec-sum}
In these proceedings, we investigated the global spin alignment of vector mesons, including $\phi$ and $J/\psi$. From measurements in Au+Au collisions at $\sqrt{s_{\mathrm{NN}}}$ = 19.6 GeV, we uncovered intriguing, strong dependency on rapidity for $\phi$ global spin alignment, shedding light on the intricate interplay between the fluctuating strong force field and meson spin alignment. The $J/\psi$ global spin alignment  with 
respect to first order event plane is found to be consistent with 1/3 within error. We further looked into the feasibility of extracting $\rho_{00}$ for $\rho^0$ meson, which can have potential impact on our understanding of the Chiral Magnetic Effect in nuclear collisions.
%
%
%

\end{document}